\documentclass[a4paper]{jpconf}
\usepackage{amsmath,amssymb,amsfonts,amsthm}
\usepackage{graphicx}

\newcommand{\be}{\begin{equation}}
\newcommand{\ee}{\end{equation}}
\newcommand{\beq}{\begin{eqnarray}}
\newcommand{\eeq}{\end{eqnarray}}

\begin{document}
\title{Highlights of Noncommutative Spectral Geometry}

\author{Mairi Sakellariadou}

\address{Theoretical Particle Physics and Cosmology Group, Department
  of Physics, King's College, University of London, Strand WC2R 2LS,
  London, U.K.}

\ead{Mairi.Sakellariadou@kcl.ac.uk}

\begin{abstract}
A summary of noncommutative spectral geometry as an approach to
unification is presented. The r\^ole of the doubling of the algebra,
the seeds of quantization and some cosmological implications are
briefly discussed.
\end{abstract}

\section{Introduction}
In the various attempts to quantize gravity, it is either assumed a
purely gravitational theory without any matter fields, or it is
considered that the interaction between gravity and matter is the most
important element of the dynamics.  I will adopt here the second
approach.

Much below the Planck energy scale, gravity can be considered as a
classical theory, and the laws of physics can be described to a good
approximation by an effective action and continuum fields. As the 
energies however approach the Planck scale, the quantum nature of space-time
becomes apparent, and the simple prescription, dictating that physics
can be described by the sum of the Einstein-Hilbert and the Standard
Model (SM) action ceases to be valid.  In the framework of
NonCommutative Spectral Geometry (NCSG)~\cite{ccm}, gravity and the
SM fields  were put together into matter and geometry on a
noncommutative space made from the product of a four-dimensional
standard commutative manifold by a noncommutative internal space.

In what follows, I will briefly present the elements of NCSG as an
approach to unification~\cite{Sakellariadou:2010nr} and highlight the
relation between the doubling of the algebra and the gauge
fields~\cite{mag}, an essential element to make the link with the
SM. I will then discuss how the doubling of the algebra is related to
dissipation, which incorporates the seeds of quantization~\cite{mag}.
I will finally discuss briefly some cosmological
consequences~\cite{Nelson:2008uy,Nelson:2009wr,mmm,Nelson:2010ru,Nelson:2010rt},
since the model lives by construction in high energy scales, offering
a natural framework to address early universe cosmology.

\section{Elements of noncommutative spectral geometry}
NCSG~\cite{ncg-book1, ncg-book2} is based on a two-sheeted space, made
  from the product of a four-dimensional smooth compact Riemannian
  manifold ${\cal M}$ (a continuous geometry for space-time), by a
  discrete noncommutative space ${\cal F}$ (an internal geometry for
  the SM) composed by only two points.  The noncommutative nature of
  the discrete space ${\cal F}$ is given by a spectral triple $({\cal
  A, H, D})$, where ${\cal A}$ is an involution of operators on the
  finite-dimensional Hilbert space ${\cal H}$ of Euclidean fermions,
  and ${\cal D}$ is a self-adjoint unbounded operator in ${\cal H}$.
  All information about space is encoded in the algebra of coordinates
  ${\cal A}$, which is related to the gauge group of local gauge
  transformations.

Assuming the algebra ${\cal A}$ to be symplectic-unitary, it
is~\cite{Chamseddine:2007ia} $\mathcal{A}=M_{a}(\mathbb{H})\oplus
M_{k}(\mathbb{C})$, with $k=2a$ and $\mathbb{H}$ denoting the algebra
of quaternions. The choice $k=4$ is the first value that produces the
correct number ($k^2=16$) of fermions in each of the three
generations, with the number of generations being a physical input.
While the choice of algebra ${\cal A}$ constitutes the main input of
the theory, the choice of Hilbert space ${\cal H}$ is irrelevant.  The
operator ${\cal D}$ corresponds to the inverse of the Euclidean
propagator of fermions, and is given by the Yukawa coupling matrix
which encodes the masses of the elementary fermions and the
Kobayashi--Maskawa mixing parameters.  The SM fermions provide the
Hilbert space ${\cal H}$ of a spectral triple for the algebra ${\cal
A}$, while the bosons are obtained through inner fluctuations of the
Dirac operator of the product geometry.

One applies the spectral action principle~\cite{spactpr}, stating that
the bare bosonic Euclidean action is the trace of the heat kernel
associated with the square of the Dirac operator and is of the form
${\rm Tr}(f({\cal D}/\Lambda))$; $f$ is a cut-off function and
$\Lambda$ fixes the energy scale. This action can be seen {\sl \`a la}
Wilson as the bare action at scale $\Lambda$.  The fermionic term can
be included by adding $(1/2)\langle J\psi,D\psi\rangle$, where $J$ is
the real structure on the spectral triple and $\psi$ is a spinor in
the Hilbert space of the quarks and leptons.  For the four-dimensional
Riemannian geometry, the trace is expressed perturbatively in terms of
the geometrical Seeley-deWitt coefficients $a_n$~\cite{sdw-coeff}:
\be\label{asymp-exp} {\rm Tr}(f(D/\Lambda))\sim
2\Lambda^4f_4a_0+2\Lambda^2f_2a_2+f_0a_4+\cdots
+\Lambda^{-2k}f_{-2k}a_{4+2k}+\cdots~.
\ee
Since its Taylor expansion at zero vanishes, it reduces to
\be {\rm Tr}(f(D/\Lambda))\sim
2\Lambda^4f_4a_0+2\Lambda^2f_2a_2+f_0a_4~;  \ee
$f$ plays a r\^ole only through its momenta $f_0, f_2, f_4$, which are
three real parameters, related to the coupling constants at
unification, the gravitational constant, and the cosmological
constant, respectively. The computation of this asymptotic expression
results to the full Lagrangian for the SM minimally coupled to
gravity, with neutrino mixing and Majorana mass terms.

This purely geometric approach to the SM leads to the correct
representations of the fermions with respect to the gauge group of the
SM, the Higgs doublet appears as part of the inner fluctuations of the
metric, and Spontaneous Symmetry Breaking mechanism arises naturally
with the negative mass term without any tuning~\cite{ccm}. The see-saw
mechanism is obtained, the 16 fundamental fermions are recovered, and
a top quark mass of $\sim 179 ~{\rm GeV}$ is predicted~\cite{ccm}.
The model also predicts the correct order of magnitude for the Higgs
mass. Strictly speaking, the predicted Higgs mass of approximately
170 GeV is ruled out from the experimental data, nevertheless it is
rather remarkable that the order of magnitude is correct, given that
the NCGS approach based on the particular choice for the algebra
${\cal A}$ must be seen as an effective theory.

\section{Dissipation and the origin of quantization}
The central ingredient in the
NCSG, namely the doubling of the algebra ${\cal A}={\cal A}_1\otimes
{\cal A}_2$ acting on the space ${\cal H} = {\cal H}_1\otimes {\cal
  H}_2$ is
related to dissipation and to the gauge
field structure~\cite{mag}.  

To highlight the justification of this claim let us consider the
equation of the classical one-dimensional damped harmonic oscillator
$m \ddot x + \gamma \dot x + k x  = 0$,
with time independent $m$, $\gamma$ and $k$, which is a simple
prototype of open systems.  In the canonical formalism for open
systems, the doubling of the degrees of freedom is required in such a
way as to complement the given open system with its time-reversed
image, thus obtaining a globally closed system for which the
Lagrangian formalism is well defined. Considering the oscillator in
the doubled $y$ coordinate $m \ddot y - \gamma \dot y + k y = 0
$ and then using the coordinates $x_{1}(t) = (x(t) + y(t))/\sqrt{2}$
and $x_{2}(t) = (x(t) - y(t))/\sqrt{2}$, the Lagrangian of this closed system
takes the form
\beq L &=& {1
  \over 2m} (m{\dot x_1} + {e_1 \over{c}} A_1)^2 - {1 \over 2m}
(m{\dot x_2} + {e_2 \over{c}} A_2)^2 - {e^2\over 2mc^2}({A_1}^2 +
{A_2}^2) - e\Phi \label{2.24i}~, \eeq
where we have introduced the vector
potential 
$A_i = (B/ 2) \epsilon_{ij} x_j$ for $i,j = 1,2$ with $B \equiv \gamma
\,{c/{e}}$ and ${\epsilon}_{ii} = 0$, ~${\epsilon}_{12} = -
{\epsilon}_{21} = 1$.  It describes two particles with opposite
charges $e_1 = - e_2 = e$ in the (oscillator) potential $\Phi \equiv
(k/2e)({x_1}^2 - {x_2}^2) \equiv {\Phi}_1 - {\Phi}_2$ with $ {\Phi}_i
\equiv (k/2/e){x_i}^{2}$ and in the constant magnetic field
$\bf{B}$ defined as $\bf{B}= \bf {\nabla} \times
\bf{A}$.

The doubled coordinate, e.g., $x_2$ acts as the gauge field component
$A_1$ to which the $x_1$ coordinate is coupled, and {\sl vice
versa}. The energy dissipated by one of the two systems is gained by
the other.  The gauge field acts as the bath or reservoir in which the
system is embedded~\cite{mag}.

The NCSG classical construction carries implicit in the doubling of
the algebra the seeds of quantization~\cite{mag}. 't~Hooft has
conjectured that, provided some specific energy conditions are met and
some constraints are imposed, loss of information might lead to a
quantum evolution~\cite{'tHooft:1999gk}.  By considering the classical
damped harmonic oscillator and its time--reversed image, we have
shown~\cite{mag} that the obtained Hamiltonian belongs to the class of
Hamiltonians considered by 't~Hooft.  We have shown~\cite{mag} that
the dissipation term in the Hamiltonian is responsible for the zero
point contribution to the energy, which is the signature of
quantization.

\section{Cosmological consequences}
In the low-energy limit the corrections to the background Einstein's
equations do not occur at the level of a
Friedmann-Lema\^{i}tre-Robertson-Walker (FLRW)
background~\cite{Nelson:2008uy}. One may have naively claimed that
this was expected, arguing that in a spatially homogeneous space-time
the spatial points are equivalent and any noncommutative effects are
then expected to vanish. However, this argument does not apply here;
the noncommutativity is incorporated in the internal manifold ${\cal
F}$ and the space-time is a commutative four-dimensional manifold.
The coupling between the Higgs field and the background geometry can
no longer be neglected once the energies reach the Higgs scale, in
which case the nonminimal coupling of Higgs field to curvature leads
to corrections to Einstein's equations even for homogeneous and
isotropic cosmological models~\cite{Nelson:2008uy}. The effect of the
nonminimal coupling of the Higgs field can be seen in two ways: it
leads to an effective gravitational constant, or it increases the
Higgs mass~\cite{Nelson:2008uy}.

The nonminimal coupling between the Higgs field and the Ricci
curvature may turn out to be crucial in early universe
cosmology~\cite{Nelson:2009wr,mmm}.  Such a coupling has been
introduced {\sl ad hoc} in the literature, in an attempt to drive
inflation through the Higgs field.
However, the value of the coupling constant between
the scalar field and the background geometry should be dictated by the
underlying theory.  Actually, even if classically the coupling between
the Higgs field and the Ricci curvature could be set equal to zero, a
nonminimal coupling will be induced once quantum corrections in the
classical field theory are considered.

We have thoroughly investigated~\cite{mmm} where the Higgs field could
play the r\^ole of the inflaton leading to a sufficient period of
inflation with induced temperature anisotropies which are in agreement
with the current measurements. The Higgs potential is
$V(H)=\lambda_0H^4-\mu_0^2H^2$, with $\mu_0$ and $\lambda_0$ subject
to radiative corrections as functions of energy.  For large enough
values of the Higgs field, the normalized value of $\mu_0$ and
$\lambda_0$ must be calculated.  We have shown~\cite{mmm} that for
each value of the top quark mass there is a value of the Higgs mass
where the effective potential is about to develop a metastable minimum
at large values of the Higgs field and the Higgs potential is locally
flattened.  Calculating~\cite{mmm} the renormalization of the Higgs
self-coupling up to two-loops, we have constructed an effective
potential which fits the renormalization group improved potential
around the flat region. There is a very good analytic fit to the Higgs
potential around the minimum of the potential, namely~\cite{mmm} :
\be
V^\text{eff}=\lambda_0^\text{eff}(H)H^4
=[a\ln^2(b\kappa H)+c] H^4~,
\ee
where the parameters $a, b$ are related to the low energy values of
top quark mass $m_{\rm t}$ as~\cite{mmm}
\beq
a(m_\text{t})&=&4.04704\times10^{-3}-4.41909\times10^{-5}
\left(\frac{m_\text{t}}{{\rm GeV}}\right)
+1.24732\times10^{-7}\left(\frac{m_\text{t}}{{\rm GeV}}\right)^2~,
\nonumber\\ 
b(m_\text{t})&=&\exp{\left[-0.979261
\left(\frac{m_\text{t}}{{\rm GeV}}-172.051\right)\right]}~.
\eeq
The parameter $c$ encodes the appearance of an extremum and depends on
the values for top quark mass and Higgs mass.
A search in the parameter space using a Monte-Carlo chain
has shown~\cite{mmm} that even
though slow-roll inflation can be realized -- a result which does not
hold for minimally coupled Higgs field -- the resulting ratio of
perturbation amplitudes is too large for any experimentally allowed
values for the masses of the top quark and the Higgs boson. 
 
Finally, by considering the energy lost to gravitational radiation by
orbiting binaries and requiring the magnitude of deviations from
General Relativity (GR) obtained within the NCSG context, to be less
than the allowed uncertainty in the data, we
imposed~\cite{Nelson:2010ru} an upper limit to the moment $f_0$, which
is used to specify the initial conditions on the gauge couplings. In
particular, by setting $\beta^2=(5\pi)/(48 Gf_0)$ we
imposed~\cite{Nelson:2010ru} the lower limit: $\beta > 7.55\times
10^{-13}~{\rm m}^{-1}$.  This observational constraint may seem weak,
however it is comparable to existing constraints on similar, {\sl ad
hoc}, additions to GR.  Moreover, since the strongest constraint comes
from systems with high orbital frequencies, future observations of
rapidly orbiting binaries, relatively close to the Earth, could
improve it by many orders of magnitude.  Thus, by purely astrophysical
observations we were able to constrain the natural length, defined
through the $f_0$ momentum of the cut-off function $f$ at which the
noncommutative effects become dominant.

\section{Conclusions}
NCSG offers an elegant and purely geometric interpretation of the SM
of electroweak and strong interactions. The doubling of the algebra is
an essential element in order to get the gauge fields of the SM.
Moreover, the NCSG classical construction carries implicit in its
feature of the doubling of the algebra the seeds of quantization.
The NCSG model lives at unification scale; thus it
provides an excellent framework to address early universe cosmological
questions, while to study astrophysical consequences one will have to
go beyond the perturbative approach.

\end{document}